\documentclass[conference]{IEEEtran}
\IEEEoverridecommandlockouts
\usepackage[utf8]{inputenc}
\usepackage{cite}
\usepackage{amsmath,amssymb,amsfonts}
\usepackage{graphicx}
\usepackage{textcomp}
\usepackage{xcolor}
\usepackage{booktabs} 
\usepackage{multirow} 
\usepackage{siunitx} 
\usepackage[hyphens]{url} 
\usepackage{hyperref}

\usepackage{algorithm}
\usepackage{algorithmic}

\usepackage{tikz}
\usetikzlibrary{shapes.geometric, arrows.meta, positioning, calc, fit}

\usepackage{balance}

\def\BibTeX{{\rm B\kern-.05em{\sc i\kern-.025em b}\kern-.08em
    T\kern-.1667em\lower.7ex\hbox{E}\kern-.125emX}}

\begin{document}

\title{Multi-view Semantic-aware Resource Allocation for Edge AI}

\author{
    \IEEEauthorblockN{Jiawei Luo$^{*}$, Hussein Saleh$^{*}$, Mohamad Assaad\thanks{$^{*}$These authors contributed equally to this work.} }
    \IEEEauthorblockA{
        \textit{Laboratoire des Signaux et Systèmes (L2S), CentraleSupélec, Université Paris-Saclay}
        \textit{Gif-sur-Yvette, France}
    }
}
\maketitle

\begin{abstract}
Semantic communication is emerging as a pivotal paradigm to overcome wireless resource scarcity in 6G networks. This paper proposes a novel end-to-end framework for semantics-aware resource allocation in multi-user Edge AI. We extend the Opportunistic Information Bottleneck (OIB) to a multi-view setting, termed the Multi-view Gaussian Information Bottleneck (MGIB),  by fusing complementary source views without variational approximation or network retraining. A resource allocation framework is then formulated for joint compression-level selection, transmission mode selection (i.e., the number of active sources per group), and PRB assignment. The  problem is formulated as a Mixed-Integer Linear Program (MILP), solved efficiently via a proposed Lagrangian Relaxation (LR) algorithm, and benchmarked against the NP-hard optimal solution. Simulations demonstrate that the proposed LR scheme achieves near-optimal utility, and doubles PRB efficiency versus average allocation.
\end{abstract}

\section{Introduction}
The sixth-generation (6G) networks are expected to support a new wave of intelligent applications, including autonomous driving, augmented and virtual reality (AR/VR), and the Internet of Things (IoT). This deep integration of communication and artificial intelligence (AI) challenges the classical Shannon paradigm \cite{b1}, which focuses purely on transmitting bits without regard to their meaning or task relevance \cite{b2}. To overcome the limitations of finite spectrum and energy, networks must shift toward goal-oriented communication (GOC) by transmitting only what is semantically meaningful for the task at hand \cite{b3}.

The Information Bottleneck (IB) principle provides a powerful framework for this setting \cite{b4}, extracting compressed representations that remain maximally informative about a target task. While the Gaussian IB (GIB) \cite{b5} admits a closed-form solution, it is restricted to Gaussian variables. General solutions like the Variational IB (VIB) \cite{b6} are computationally heavy and difficult to use in real-time systems. The Opportunistic IB (OIB) \cite{b7} bridges this gap by identifying a Gaussian layer inside a pre-trained network, enabling tractable compression without retraining. However, OIB handles only a single view, while many practical scenarios involve multiple views of the same scene.

Multi-view learning aims to leverage complementary information from multiple representations of the same data. Classical methods such as DCCA \cite{b8} learn shared cross-view representations through deep canonical correlation, but are not designed for task-driven or variable-rate compression. Works like DMIB \cite{b9} and MEIB \cite{b10} extend the IB principle to the multi-view domain, explicitly optimizing the trade-off between compression and task relevance across views. However, they rely on variational lower bounds to approximate the intractable mutual information terms, which introduces computational overhead and does not guarantee bound tightness. TMC \cite{b11} and RCML \cite{b12} achieve strong multi-view classification performance through self-supervised and reliability-aware mechanisms, but similarly lack a principled compression-rate control tied to physical resources.

This paper addresses these limitations by extending OIB~\cite{b7} to the multi-view setting and using the resulting framework to design an efficient resource allocation system for multi-user Edge AI. By identifying a Gaussian layer within a pre-trained network for each view, MGIB applies the closed-form GIB solution~\cite{b5} to construct view-specific compression matrices analytically. This retains full computational tractability while enabling state-of-the-art fusion performance --- without any variational approximation or network retraining. Our contributions are:
\begin{itemize}
    \item We extend the OIB framework to the multi-view setting, enabling the fusion of complementary information across multiple views while preserving the closed-form tractability of GIB.
    \item We develop a system model linking MGIB's tunable compression dimension ($n_z$) to Physical Resource Block (PRB) costs in a multi-user wireless uplink, and formulate the joint compression,
    mode selection (i.e., the number of sources selected for transmission), and PRB assignment as a Mixed-Integer Linear
    Program (MILP).
    \item We propose a Lagrangian Relaxation (LR)-based algorithm  which exploits the separable structure of the MILP to decompose the problem into $K$ independent per-group subproblems, each solved by a channel-aware greedy procedure. The LR achieves near-optimal utility across all PRB regimes at a fraction of the computational cost of the MILP benchmark.
\end{itemize}
\section{Preliminaries}

\subsection{Information Bottleneck ($IB$)}
 
The IB principle~\cite{b4} formalizes the trade-off between compressing a source $X$ into a representation $Z$ and preserving
its relevance for a target $Y$, via the Lagrangian:
\begin{equation}
    \mathcal{L}_{\mathrm{IB}} = I(X;Z) - \beta\, I(Z;Y)
\end{equation}
where $I(.;.)$ is the mutual information, and $\beta > 0$ controls the compression-relevance trade-off. Solving this problem in general requires intractable mutual information estimation. General solutions such as the Variational IB (VIB)~\cite{b6} approximate these terms via variational bounds but introduce significant computational overhead, making them unsuitable for real-time systems. This motivates the Gaussian special case described next.

\subsection{Gaussian Information Bottleneck ($GIB$)}
 
When $(X,Y) \sim \mathcal{N}(0, \Sigma_{XY})$, the IB problem admits a closed-form solution~\cite{b5}. The optimal compression mapping is linear:
\begin{equation}
    Z = A_\rho X + \xi, \quad \xi \sim \mathcal{N}(0,I)
\end{equation}
where $\xi$ is an independent additive Gaussian noise vector. The compression matrix $A_\rho$ is constructed from the eigen-decomposition of $\Sigma_{X|Y}\Sigma_X^{-1}$, yielding eigenvalues $\{\lambda_i\}$ and eigenvectors $\{v_i\}$, sorted in ascending order. Two key quantities are defined~\cite{b5}:
\begin{itemize}
    \item \textit{Critical threshold:} $\beta_i^c = \frac{1}{1-\lambda_i}$,
          determining when feature $i$ is activated.
    \item \textit{Loading coefficient:} $\alpha_i =
          \sqrt{\frac{\beta(1-\lambda_i)-1}{\lambda_i r_i}}$,
          where $r_i = v_i^T \Sigma_X v_i$.
\end{itemize}
The full matrix $A$ is constructed as a piecewise function of $\beta$:
\begin{equation}
A = \begin{cases}
[0^T, \ldots, 0^T], & 0 < \beta \leq \beta_1^c \\
[\alpha_1 v_1^T, 0^T, \ldots, 0^T], & \beta_1^c < \beta \leq \beta_2^c \\
\vdots \\
[\alpha_1 v_1^T, \ldots, \alpha_{n_x} v_{n_x}^T], & \beta > \beta_{n_x}^c
\end{cases}
\end{equation}
The final compression matrix $A_\rho$ consists of the non-zero rows of $A$. As $\beta$ increases past each threshold $\beta_i^c$, a new row is added to $A_\rho$, increasing the number of extracted features $n_z$. This provides a direct, tunable compression mechanism without any variational approximation or network retraining.

\subsection{Opportunistic Information Bottleneck ($OIB$)}
 
The OIB framework~\cite{b7} extends GIB to practical deep learning settings by identifying an intermediate layer within a pre-trained network where the input-output relationship is approximately Gaussian, typically after a DFT transformation. By defining a linear regression sub-task at this layer, OIB applies the closed-form GIB solution to design a tunable feature extractor without retraining the entire network. The compression matrix $A_\rho$ is computed analytically and $n_z$ remains directly controllable via $\beta$. However, OIB is confined to a single input view, which motivates our multi-view extension presented in the next section.

\section{System Model and the Optimization Problem}
\begin{figure*}[t]
\centering
\begin{tikzpicture}[
    scale=0.7, transform shape, font=\small,
    node distance=1.2cm and 1cm,
    block/.style={rectangle, draw, fill=blue!10, text width=7em, text centered, rounded corners, minimum height=3.2em},
    smallblock/.style={rectangle, draw, fill=green!10, text width=5.5em, text centered, rounded corners, minimum height=2.8em},
    sum/.style={circle, draw, fill=red!10, minimum size=1em},
    line/.style={draw, -{Latex[length=2mm]}},
    dashedline/.style={draw, dashed, -{Latex[length=2mm]}}
]
% User Equipment (UE) Side Nodes
\node (x1) [text width=8em] {Raw Data $\mathbf{x}$ (View 1)};
\node (x2) [below=2.2cm of x1, text width=8em] {Raw Data $\mathbf{x}$ (View 2)};
\node (dft1) [smallblock, right=of x1] {DFT $\tilde{\mathbf{x}}^{(1)}$};
\node (dft2) [smallblock, right=of x2] {DFT $\tilde{\mathbf{x}}^{(2)}$};
\node (comp1) [block, right=of dft1] {MGIB Compression $\mathbf{z}^{(1)}=\mathbf{A}^{(1)}\tilde{\mathbf{x}}^{(1)}$};
\node (comp2) [block, right=of dft2] {MGIB Compression $\mathbf{z}^{(2)}=\mathbf{A}^{(2)}\tilde{\mathbf{x}}^{(2)}$};

% Wireless Channel
\node (sum1) [sum, right=1.5cm of comp1] {$+$};
\node (sum2) [sum, right=1.5cm of comp2] {$+$};
\node (noise1) [below=0.3cm of sum1] {Noise $\mathbf{n}$};

% Edge Server (ES) Side Nodes
\node (recon1) [block, right=1.5cm of sum1] {Reconstruction $\mathbf{\hat{a}}_0^{(1)}$};
\node (recon2) [block, right=1.5cm of sum2] {Reconstruction $\mathbf{\hat{a}}_0^{(2)}$};
\node (fusion) [block, fill=orange!20, right=of recon1, yshift=-1.8cm] {Attention Fusion};
\node (classifier) [block, fill=purple!20, right=of fusion] {Classifier $C(\cdot)$};
\node (output) [right=of classifier] {Label $\hat{y}$};

% Connections
\draw[line] (x1) -- (dft1);
\draw[line] (x2) -- (dft2);
\draw[line] (dft1) -- (comp1);
\draw[line] (dft2) -- (comp2);
\draw[line] (comp1) -- (sum1);
\draw[line] (comp2) -- (sum2);
\draw[dashedline] (noise1) -- (sum1);
\draw[dashedline] (noise1) -- (sum2);
\draw[line] (sum1) -- (recon1);
\draw[line] (sum2) -- (recon2);
\draw[line] (recon1) -- (fusion.west |- recon1.east);
\draw[line] (recon2) -- (fusion.west |- recon2.east);
\draw[line] (fusion) -- (classifier);
\draw[line] (classifier) -- (output);

% --- Dashed bounding boxes and labels ---
\node [draw, dashed, rounded corners, inner sep=10pt, fit=(x1) (x2) (dft1) (dft2) (comp1) (comp2), label={[font=\bfseries]above:User Equipment (UE)}] {};
\node [draw, dashed, rounded corners, inner sep=10pt, fit=(recon1) (recon2) (fusion) (classifier) (output), label={[font=\bfseries]above:Edge Server (ES)}] {};

\end{tikzpicture}
\caption{End-to-end MGIB-based semantic communication system for multi-view Edge AI inference.}
\label{fig:system_diagram}
\end{figure*}
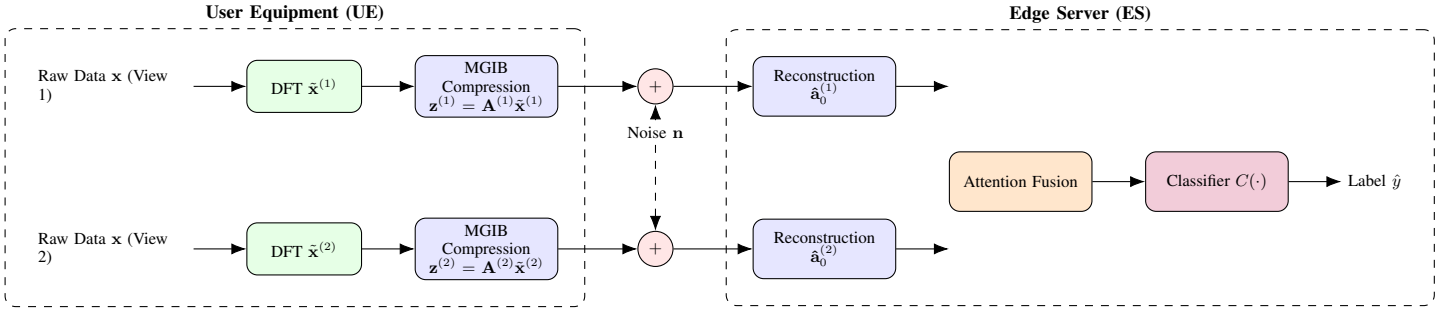

We now describe the system model and formulate the resource allocation problem. A workflow diagram of the proposed end-to-end system is shown in Fig.~\ref{fig:system_diagram}.

We consider a multi-user wireless uplink system with $K$ groups, each consisting of $C$ sources that capture multiple views of the same physical process. A total of $R$ Physical Resource Blocks (PRBs) are available for allocation across all sources. Given the scarcity of PRB resources relative to the number of groups, transmitting raw data from all sources is infeasible. To maximize inference accuracy under this constraint, we leverage the complementarity between source views within each group --- rather than transmitting redundant single-view data, each group fuses information from multiple views to extract a richer and more compact semantic representation. This multi-view fusion is realized through our proposed Multi-view Gaussian Information Bottleneck (MGIB) framework.

\subsection{MGIB Framework}

The MGIB framework extends OIB~\cite{b7} to the multi-view setting. The key insight is that for each view $v$, an approximately Gaussian layer can be identified within a shared pre-trained network after a DFT transformation. Applying the GIB solution independently to each view yields view-specific compression and reconstruction matrices $\{A_{n_z}^{(v)}, \Theta_{n_z}^{(v)}\}$ for a range of feature dimensions $n_z$, governed by the trade-off parameter $\beta$.

\subsubsection{Offline Stage}
The offline process constructs the view-specific feature extractors following a five-step pipeline summarized in Algorithm~\ref{alg:mgib_setup}. The key steps are: (1) data preparation and DFT transformation; (2) training a reference model $M_{\mathrm{ref}}$; (3) generating view-specific regression targets; (4) solving the GIB generalized eigenvalue problem per view to obtain $\{\lambda_i, v_i\}$; and (5) computing and storing $A_{n_z}^{(v)}$ and $\Theta_{n_z}^{(v)}$ for each desired $n_z$. An attention module $M_{\mathrm{attn}}$ is then trained on the reconstructed features with the classifier frozen.

\subsubsection{Online Stage}
During inference, each source performs efficient linear compression of its DFT-transformed input $\tilde{x}^{(v)}$:
\begin{equation}
    z^{(v)} = A_{n_z}^{(v)}\, \tilde{x}^{(v)}
\end{equation}
The compressed features are transmitted over the wireless channel to the edge server, which performs L-MMSE reconstruction:
\begin{equation}
    \hat{a}_0^{(v)} = \mathrm{ReLU}\!\left(\Theta_{n_z}^{(v)}\, z^{(v)}\right)
\end{equation}
The reconstructed features from both views are then fused via the learned attention module, which generates an element-wise map $m$:
\begin{equation}
    a_{\mathrm{fused}} = m \odot \hat{a}_0^{(1)} + (1-m) \odot \hat{a}_0^{(2)}
\end{equation}
The fused features are passed to the classifier $C(\cdot)$ for final inference. This architecture allows the edge server to adaptively weight the two views based on their quality and complementarity, achieving state-of-the-art fusion performance while retaining the closed-form tractability of GIB.

\begin{algorithm}[t]
\caption{Offline Setup for MGIB Framework}
\label{alg:mgib_setup}
\begin{algorithmic}[1]
\STATE \textbf{Input:} Raw training data $\mathcal{D}_{\text{train}}$, view transformations $T_1, T_2$.
\STATE \textbf{Output:} Trained $M_{\text{attn}}$, classifier $C(\cdot)$, matrix sets $\{\boldsymbol{A}_{n_z}^{(v)}\}, \{\boldsymbol{\Theta}_{n_z}^{(v)}\}$.
\STATE \textit{Data and Model Preparation}
\STATE $\boldsymbol{X}_{\text{orig}} \leftarrow \text{DFT}(\mathcal{D}_{\text{train}})$
\STATE Train $M_{\text{ref}}(\cdot) = C(\text{ReLU}(\text{L0}_{\text{ref}}(\cdot)))$ on $\boldsymbol{X}_{\text{orig}}$.
\STATE Freeze weights of the trained $M_{\text{ref}}$.
\STATE \textit{View-Specific OIB Matrix Generation}
\FOR{$v = 1$ to $2$}
    \STATE $\boldsymbol{X}^{(v)} \leftarrow \text{DFT}(T_v(\mathcal{D}_{\text{train}}))$
    \STATE $\boldsymbol{Y}_{\text{tilde}}^{(v)} \leftarrow \text{L0}_{\text{ref}}(\boldsymbol{X}^{(v)}) + \boldsymbol{\epsilon}$
    \STATE Solve GEP using $(\boldsymbol{X}^{(v)}, \boldsymbol{Y}_{\text{tilde}}^{(v)})$.
    \FOR{each desired dimension $n_z$}
        \STATE Compute and store $\boldsymbol{A}_{n_z}^{(v)}$ and $\boldsymbol{\Theta}_{n_z}^{(v)}$.
    \ENDFOR
\ENDFOR
\STATE \textit{Attention Module Training}
\STATE Select a representative dimension $n_z^*$.
\STATE Generate reconstructed features $\boldsymbol{\hat{a}}_{0}^{(1)}, \boldsymbol{\hat{a}}_{0}^{(2)}$ using $\boldsymbol{A}_{n_z^*}^{(v)}, \boldsymbol{\Theta}_{n_z^*}^{(v)}$.
\STATE Train $M_{\text{attn}}$ on $(\boldsymbol{\hat{a}}_{0}^{(1)}, \boldsymbol{\hat{a}}_{0}^{(2)})$ with frozen classifier $C(\cdot)$.
\end{algorithmic}
\end{algorithm}

\subsection{Resource Optimization Problem Formulation}
We now formulate the  resource allocation problem. For each group $k$, three decisions are jointly optimized: (i) the compression level $n_z$, controlling how many semantic features are extracted and transmitted; (ii) the transmission mode $j$, denoting the number of active sources selected for transmission; and (iii) the PRB assignment, mapping each active source to a subset of the available spectrum. These decisions are tightly coupled: the compression level determines the required transmission rate, the mode determines how many sources need PRBs, and the channel quality of assigned PRBs determines whether the rate constraint can be met.

To encode these decisions, we define a single binary variable $v_{n_z,k,j} \in \{0,1\}$ jointly capturing the compression level and transmission mode, along with $a_{k,c} \in \{0,1\}$ and $x_{k,c,r} \in \{0,1\}$ indicating whether source $(k,c)$ is active and whether PRB $r$ is assigned to it, respectively. The network-wide problem is to maximize total utility:
\begin{equation}
  \max \sum_{k} \sum_{n_z} \sum_{j} v_{n_z,k,j} \cdot U(n_z,j)
\end{equation}
subject to:
\begin{gather}
  \sum_{n_z,j} v_{n_z,k,j} \leq 1 \quad \forall\, k \tag{C1}\\
  \sum_{c} a_{k,c} = \sum_{n_z,j} m_j \cdot v_{n_z,k,j} \quad \forall\, k \tag{C2}\\
  x_{k,c,r} \leq a_{k,c} \quad \forall\, k,c,r \tag{C3}\\
  \sum_{k}\sum_{c} x_{k,c,r} \leq 1 \quad \forall\, r \tag{C4}\\
  \sum_{n_z,j} n_z \cdot b \cdot v_{n_z,k,j} \leq T \cdot B_{\mathrm{RB}} \sum_{r} x_{k,c,r} \log_2(1+\gamma_{k,c,r}) \notag \\
    + M(1-a_{k,c})  \hfill\quad \forall\, k,c \tag{C5}\\
  v_{n_z,k,j},\; a_{k,c},\; x_{k,c,r} \in \{0,1\} \notag
\end{gather}
where $m_j \in \{0,1,2\}$ is the number of sources required by mode $j$ and $M$ is the Big-M constant. The Shannon capacity available to source $(k,c)$ on PRB $r$ is $C_{k,c,r} = T \cdot B_{\text{RB}} \cdot \log_2(1+\gamma_{k,c,r})$ bits, where $T$ is the slot duration, $B_{\text{RB}}$ is the PRB bandwidth, and $\gamma_{k,c,r}$ is the received SNR. Each group must transmit $n_z \cdot b$ bits per active source, where $b=16$ bits per feature dimension. Constraints C1--C5 are interpreted as follows: \textbf{C1} ensures each group selects at most one $(n_z,j)$ pair; \textbf{C2} matches the number of active sources to the selected mode; \textbf{C3} allows PRB assignment only to active sources; \textbf{C4} enforces PRB exclusivity across all sources; \textbf{C5} ensures the selected compression level is supported by the assigned PRBs, with Big-$M$ deactivating the constraint for inactive sources.

\section{Semantics-Aware Resource Allocation - Proposed Solution}

We propose a Lagrangian Relaxation (LR) approach to solve the formulated resource allocation problem efficiently. The key observation is that constraint C4 is the only constraint coupling all $K$ groups. Relaxing it allows the problem to decompose into $K$ independent subproblems, one per group, which can be solved efficiently via a greedy procedure.

\subsubsection{Relaxation}

We relax C4 by introducing a dual variable $\lambda_r \geq 0$ for each PRB $r$, interpreted as the \emph{price} of that PRB. The Lagrangian is:
\begin{equation}
    \mathcal{L}(\boldsymbol{\lambda}) = \sum_{r=1}^R \lambda_r
    + \sum_{k=1}^K \mathcal{L}_k(\boldsymbol{\lambda})
\end{equation}
where each group subproblem is:
\begin{equation}
    \mathcal{L}_k(\boldsymbol{\lambda}) = \max \sum_{n_z,j}
    v_{n_z,k,j}\, U(n_z,j)
    - \sum_{c=1}^C \sum_{r=1}^R \lambda_r\, x_{k,c,r}
\end{equation}
subject to C1, C2, C3, C5 for group $k$ only. Each group maximizes its semantic utility minus the total price paid for its assigned PRBs. The $K$ subproblems are solved independently via a channel-aware greedy procedure: for each candidate $(n_z, j)$ pair, PRBs are sorted by their bits-per-price ratio $c_{k,c,r}/(\lambda_r + \epsilon)$ and greedily assigned to each source until C5 is satisfied. The pair yielding the highest net utility is selected.

\subsubsection{Dual Update}

The dual variables are updated via the subgradient method. The subgradient of $\mathcal{L}(\boldsymbol{\lambda})$ with respect to $\lambda_r$ is:
\begin{equation}
    g_r = \sum_{k,c} x_{k,c,r} - 1
\end{equation}
which measures the overdemand for PRB $r$: positive when overloaded, negative when unused. Prices are updated as:
\begin{equation}
    \lambda_r^{(t+1)} = \left[\lambda_r^{(t)}
    + \alpha_t \cdot g_r^{(t)}\right]^+
\end{equation}
where $[\cdot]^+ = \max(0,\cdot)$ enforces non-negativity. The step size uses an aggressive diminishing rule for the first 50 iterations to spread groups quickly, followed by the Polyak formula:
\begin{equation}
    \alpha_t = \frac{\rho\left(\mathcal{L}(\boldsymbol{\lambda}^{(t)})
    - f^*_{\mathrm{best}}\right)}{\|g^{(t)}\|^2}
\end{equation}
$f^*_{\mathrm{best}}$ is the best feasible utility found so far
and $\rho \in (0,2)$.

\subsubsection{Rounding}

Since the subgradient rarely yields a perfectly conflict-free solution, a two-pass rounding procedure is applied to the final iteration's solution. In \textbf{Pass 1}, each contested PRB is awarded to the group with the highest utility and removed from all others, enforcing C4. In \textbf{Pass 2}, groups that lost PRBs search the unused PRB pool --- sorted by channel quality --- to repair their C5 constraint. If insufficient unused PRBs exist, the compression level $n_z$ is reduced to the highest value supportable by the available bits. Groups that cannot achieve feasibility under any $n_z$ are set to $j=0$ (no transmission). The full procedure is summarized in Algorithm~\ref{alg:LR}.

\begin{algorithm}
\caption{Lagrangian Relaxation for Resource Allocation}
\label{alg:LR}
\begin{algorithmic}[1]
 
\STATE \textbf{Input:} $\{c_{k,c,r}\}$, $U$, $K,C,J,N_z,R$,
       $T_{\max}$, $\rho$.
\STATE \textbf{Output:} $\{x_{k,c,r}^*,n_{z,k}^*,j_k^*\}$,
       $\{U_k^*\}$.
\STATE Initialize $\lambda_r \leftarrow 0\;\forall r$,\;
       $f^*_{\mathrm{best}} \leftarrow 0$
 
\FOR{$t = 1$ to $T_{\max}$}
    \FOR{each group $k$}
        \STATE For each $(n_z,j)$ with $j{>}0$: sort PRBs by
               $c_{k,c,r}/(\lambda_r{+}\epsilon)$; greedily
               assign to sources until C5 met; \textbf{break}
               if infeasible
        \STATE $j_k^*,n_{z,k}^* \leftarrow
               \arg\max\{U(n_z,j) -
               \textstyle\sum_{c,r}\lambda_r x_{k,c,r}\}^+$
    \ENDFOR
    \STATE $g_r \leftarrow \sum_{k,c} x_{k,c,r} - 1
           \quad\forall r$ \hfill\textit{(subgradient)}
    \IF{$\sum_r\mathbf{1}[g_r{>}0]\leq 3$ \AND
        $\sum_k U_k > f^*_{\mathrm{best}}$}
        \STATE $f^*_{\mathrm{best}}\leftarrow\sum_k U_k$;\;
               save allocation
    \ENDIF
    \IF{$\|g\|^2 < \epsilon$}
    \STATE \textbf{break}
    \ENDIF
    \STATE $\alpha_t \leftarrow 2/\sqrt{t}$ if $t{\leq}50$,
           else $\rho(\mathcal{L}(\lambda^{(t)}){-}f^*_{\mathrm{best}})/
           \|g^{(t)}\|^2$
    \STATE $\lambda_r \leftarrow [\lambda_r{+}\alpha_t g_r]^+
           \;\forall r$;\; reset if stuck every 100 iters
\ENDFOR
 
\STATE \textbf{Pass 1:} For each contested PRB $r$, keep
       highest-utility group; set $x_{k,c,r}{=}0$ for others
\STATE \textbf{Pass 2:} For groups failing C5, assign unused
       PRBs by $c_{k,c,r}\!\downarrow$; reduce $n_z$ if needed;
       set $j_k^*{=}0$ if infeasible
\FOR{each group $k$}
    \STATE $U_k^* \leftarrow U(n_{z,k}^*,j_k^*)$ if C5 met;
           else $0$
\ENDFOR
\RETURN $\{x_{k,c,r}^*,n_{z,k}^*,j_k^*\}$, $\{U_k^*\}$
 
\end{algorithmic}
\end{algorithm}
\section{Experimental Results and Analysis}
We use the MNIST dataset as a proof-of-concept benchmark, allowing a clear and interpretable evaluation of the interplay between MGIB and the resource allocation algorithm. It serves as a standard benchmark for multi-view IB methods, ensuring a fair comparison against key works like DMIB \cite{b9} and MEIB \cite{b10}. Network architectures and simulation parameters are in Table \ref{tab:nn_arch} and \ref{tab:sim_params}. For the resource allocation evaluation, we consider a multi-user wireless uplink with $K=15$ groups and $C=4$ sources per group. The PRB budget is swept over $R \in \{5, 8, 9, 10, 15, 20, 25, 30, 35, 40, 50, 60, 70\}$ to evaluate performance across severely constrained to resource-abundant regimes. Results are averaged over 500 Monte Carlo runs, each with independent Rayleigh fading channel realizations, and reported as mean $\pm$ standard deviation.

\begin{table}[h]
\centering
\vspace{-8pt}
\caption{Network Architectures}
\vspace{-7pt}
\label{tab:nn_arch}
\resizebox{\columnwidth}{!}{%
\begin{tabular}{@{}lll@{}}
\toprule
\textbf{Component} & \textbf{Layer Type} & \textbf{Shape (In $\rightarrow$ Out)} \\ \midrule
\multirow{4}{*}{\textit{M\textsubscript{ref}} Classifier} 
 & Linear + ReLU & 256 $\rightarrow$ 128 \\
 & Linear + ReLU & 128 $\rightarrow$ 64 \\
 & Linear + ReLU & 64 $\rightarrow$ 16 \\
 & Linear & 16 $\rightarrow$ 10 \\ \midrule
\multirow{3}{*}{Attention Module} 
 & Linear + ReLU & 512 (256+256) $\rightarrow$ 128 \\
 & Linear + Sigmoid & 128 $\rightarrow$ 256 \\
 & Dropout & p = 0.3 \\
 \bottomrule
\end{tabular}}
\end{table}

\begin{figure}[t]
\centering
\includegraphics[width=0.9\columnwidth]{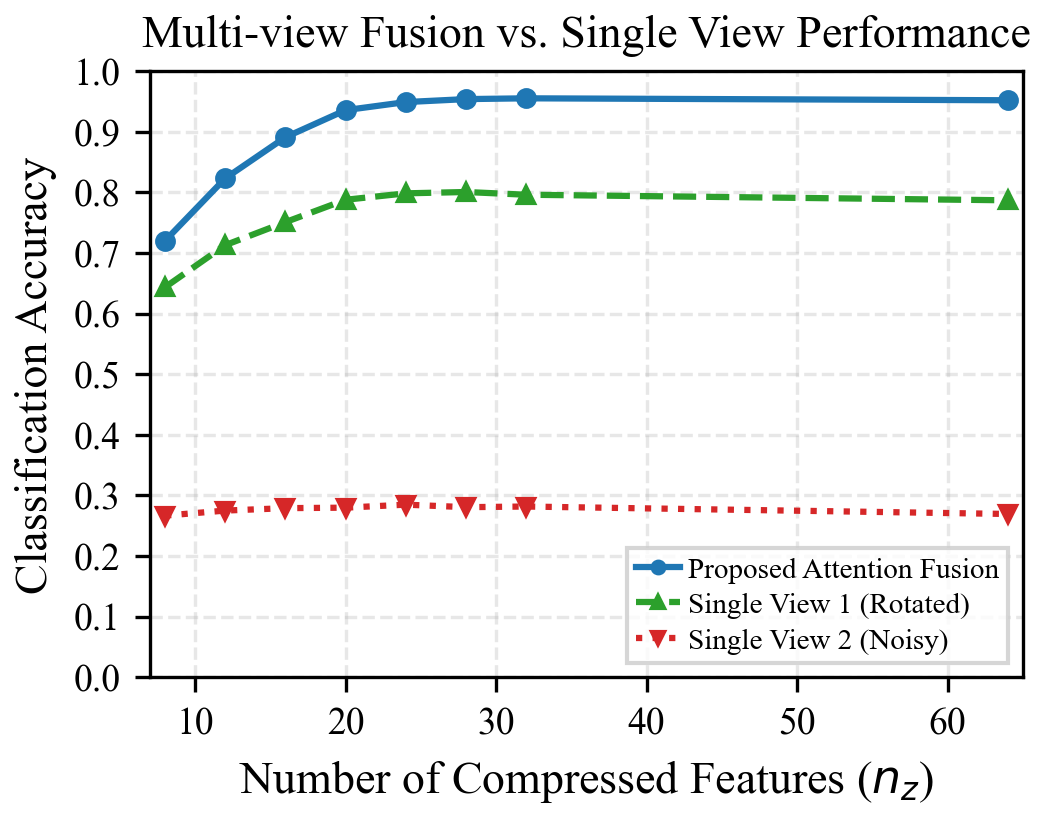}
\caption{MGIB vs. single-view baselines on MNIST Rotation+Noisy.}
\label{fig:mgib_performance}
\end{figure}

%\subsection{MGIB Performance}
First, we validate the MGIB module itself. Following the setup in \cite{b9, b10}, we create a rotated view and a noisy view of MNIST. Fig. \ref{fig:mgib_performance} shows the clear advantage of our fusion approach over single views. Table \ref{tab:sota} confirms that our MGIB approach achieves state-of-the-art accuracy, outperforming established methods. This validates MGIB as a highly effective and efficient feature extractor and fusion mechanism, laying a strong foundation for the system-level analysis.

\begin{table}[t]
\centering
\vspace{-12pt}
\caption{SOTA Comparison on MNIST Rotation+Noisy Dataset}
\vspace{-7pt}
\label{tab:sota}
\begin{tabular}{@{}lc@{}}
\toprule
\textbf{Method} & \textbf{Accuracy} \\ \midrule
Linear CCA & 0.579 \\
DCCA \cite{b8} & 0.747 \\
DCCAE & 0.622 \\
Deep Multi-view IB (SIAM 2019) \cite{b9} & 0.823 \\
MEIB (ICASSP 2022) \cite{b10} & 0.828 \\
\textbf{MGIB (ours)} & \textbf{0.950} \\ \bottomrule
\end{tabular}
\end{table}

\begin{figure}[t]
\centering
\includegraphics[width=0.9\columnwidth]{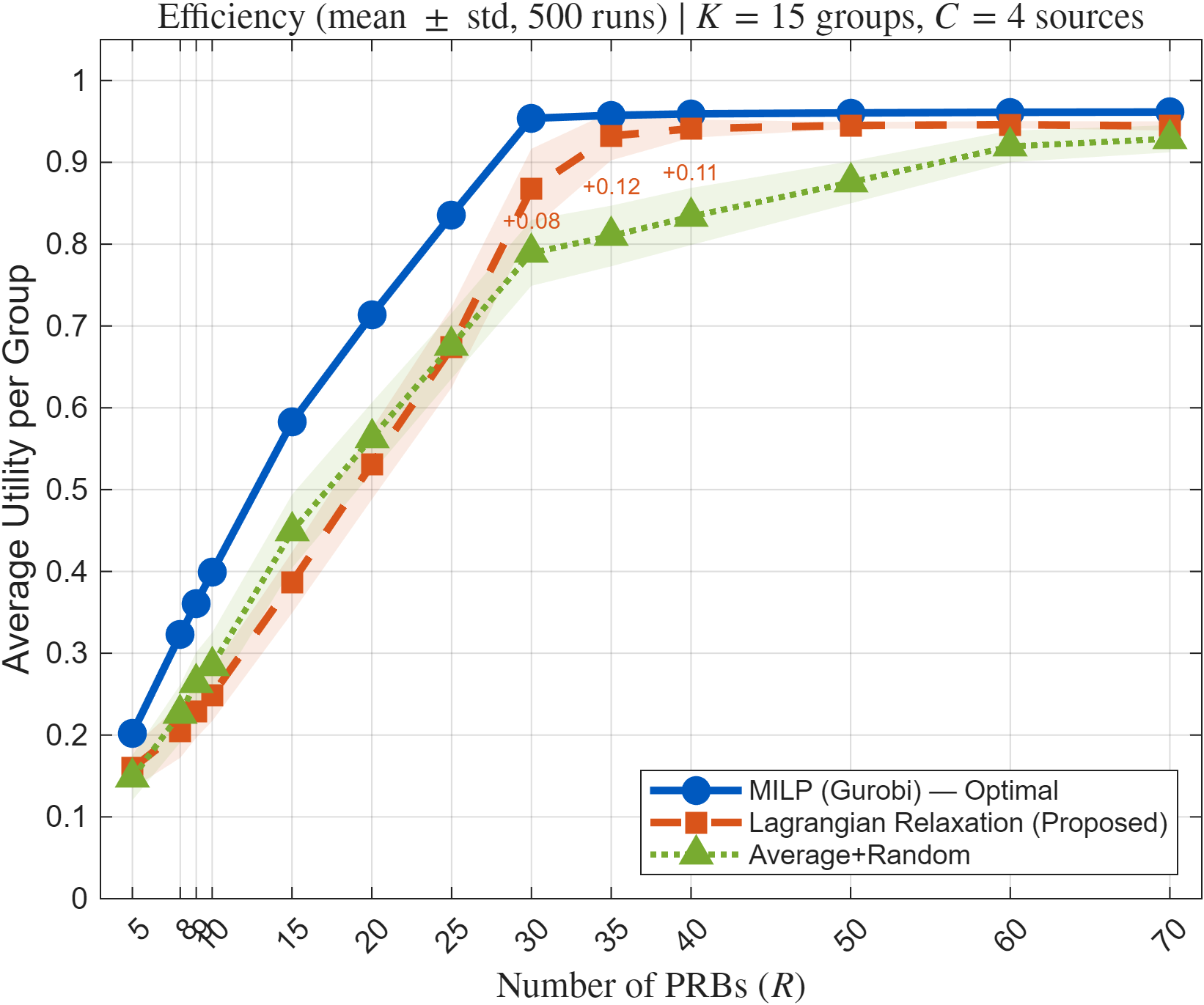}
\caption{Utility of the LR vs. MILP optimal and Average+Random baseline.}
\vspace{-15pt}
\label{fig:system_performance}
\end{figure}

\begin{table}[b] 
\centering
\vspace{-15pt}
\caption{Simulation Parameters}
\vspace{-7pt}
\label{tab:sim_params}
\begin{tabular}{@{}lc@{}}
\toprule
\textbf{Parameter} & \textbf{Value} \\ 
\midrule
Number of Groups, $K$ & \ 15 \\
Sources per Group $C$ & \ 4 \\
PRBs $R$ & $\{5, 8, 9, 10, 15, 20, 25, 30, 35, 40, 50, 60, 70\}$ \\
Cell radius $R_{\text{cell}}$ & 1 km \\
Transmit power $p$ per PRB          & 250 mW \\
Noise power $\sigma^2$        & $-104$ dBm \\
Slot duration $T$             & 0.5 ms \\
PRB bandwidth $B_{\text{RB}}$ & 180 kHz \\
Bits per feature $b$          & 16 \\
Semantic dimensions $n_z$     & \{8,10,12,14,18,22,26,32,64\} \\
\hline
\end{tabular}
\end{table}

Fig.~\ref{fig:system_performance} compares the proposed LR scheme against the MILP optimal benchmark, solved via Gurobi~\cite{gurobi} (which has high complexity since MILP is generally NP-hard), and the Average+Random baseline, in which the total PRB budget $R$ is divided equally among all $K$ groups and assigned randomly, over 500 Monte Carlo runs with independent Rayleigh fading realizations. The proposed LR scheme achieves near-optimal utility across the full PRB sweep, closely tracking the MILP upper bound. As $R$ increases, the LR rapidly diverges from Average+Random: at $R = 30$, it outperforms the baseline by $+0.08$, with the gap widening to $+0.12$ at $R = 35$ before narrowing as both schemes saturate beyond $R = 50$. The LR reaches $\approx 0.95$ utility at $R = 35$ PRBs, while Average+Random requires $R = 60$ to reach the same level --- a \textbf{$2\times$ PRB efficiency advantage}.

\section{Conclusion}
In this paper, we proposed a semantics-aware resource allocation framework for multi-user Edge AI systems. We extend the OIB framework to the multi-view setting while preserving closed-form tractability of GIB. Building on this, we formulated the joint compression, mode selection, and PRB assignment as a MILP, and proposed a Lagrangian Relaxation algorithm that decomposes the problem into efficient per-group subproblems. Simulations validated that the proposed LR scheme achieves near-optimal utility, with a $2\times$ PRB efficiency advantage over the Average+Random baseline. 

\section*{Acknowledgments}
This research was funded in part by the Agence Nationale de la Recherche (ANR) under the ANR-24-IAS1-0002-02 COMSEMA project.

\end{document}